\documentclass[vecphys]{svmult}

\usepackage{graphicx}        
\usepackage{multicol}        
\usepackage[bottom]{footmisc}

\begin{document}

\newcommand{\sun}{\mbox{$_{\odot}$}}
\newcommand{\apj}{ApJ}
\newcommand{\apjs}{ApJS}
\newcommand{\aap}{A\&A}
\newcommand{\aaps}{A\&AS}
\newcommand{\pasp}{PASP}
\newcommand{\pasj}{PASJ}
\newcommand{\aj}{AJ}
\newcommand{\qjras}{QJRAS}
\newcommand{\arcmin}{\mbox{$^{\prime}$}}
\newcommand{\arcsec}{\mbox{$^{\prime \prime}$}}
\newcommand{\iras}{{\bf{\sl IRAS}}}
\newcommand{\mic}{\mbox{$\mu{\rm m}$}}
\newcommand{\etal}{et al.}
\newcommand{\smpy}{\mbox{M$_{\odot}$ yr$^{-1}$}}
\newcommand{\xten}[1]{\mbox{$\times 10^{#1}$}}
\newcommand{\filter}[1]{\bf #1}
\newcommand{\magap}[1]{ #1$^{\mbox m}$}
\newcommand{\rf}{\par\noindent\hangindent 15pt {}}
\newcommand{\ltappeq}{\raisebox{-0.6ex}{$\,\stackrel
{\raisebox{-.2ex}{$\textstyle <$}}{\sim}\,$}}
\newcommand{\gtappeq}{\raisebox{-0.6ex}{$\,\stackrel
{\raisebox{-.2ex}{$\textstyle >$}}{\sim}\,$}}
\newcommand{\mn}{MNRAS}
\newcommand{\apspsc}{Ap\&SS}
\newcommand{\annrev}{ARA\&A}
\newcommand{\aandasupp}{A\&AS}
\newcommand{\aanda}{A\&A}
\newcommand{\aandarev}{A\&AR}
\newcommand{\nature}{Nat}
\newcommand{\icarus}{Icarus}
\newcommand{\liege}{Mem. Soc. R. Li\`{e}ge}
\newcommand{\jqsrt}{J. Quant. Spectrosc. Radiat. Transfer}
\newcommand{\soviet}{SvA}
\newcommand{\revmex}{Rev. Mex. Astron. Astrofis.}
\newcommand{\sarcsec}{\raisebox{-0.6ex}{$\,\stackrel
{\raisebox{-.2ex}{${\prime \prime}$}}{.}\,$}}
\newcommand{\half}{\mbox{$\frac{1}{2}$}}
\newcommand{\third}{\mbox{$\frac{1}{3}$}}
\newcommand{\qbar}{\mbox{$\stackrel{-}{\textstyle Q}^{i}(\nu)$}}
\newcommand{\btilde}{\mbox{$\stackrel{-}{\textstyle b}$}}
\newcommand{\mdot}{\mbox{$\stackrel{.}{M}$}}
\newcommand{\mdotw}{\mbox{$\stackrel{.}{M}_{\mbox{w}}$}}
\newcommand{\mdota}{\mbox{$\stackrel{.}{M}_{\mbox{acc}}$}}
\newcommand{\bbar}{\mbox{$\stackrel{-}{\textstyle B}_{\nu}(r)$}}
\newcommand{\gff}{\mbox{G45.13 +0.14A}}
\newcommand{\gtn}{\mbox{G29.96 $-$0.02}}
\newcommand{\bd}{\mbox{BD +30\degs 3639}}
\newcommand{\degs}{\mbox{$^{\mbox{o}}$}}
\newcommand{\teff}{\mbox{T$_{\mbox{eff}}$}}
\newcommand{\te}{\mbox{T$_{\mbox{e}}$}}
\newcommand{\nh}{\mbox{N$_{\mbox{H}}$}}
\newcommand{\kms}{\mbox{km s$^{-1}$}}
\newcommand{\ha}{\mbox{H$\alpha$}}
\newcommand{\bra}{\mbox{Br$\alpha$}}
\newcommand{\brg}{\mbox{Br$\gamma$}}
\newcommand{\pfg}{\mbox{Pf$\gamma$}}
\newcommand{\pfb}{\mbox{Pf$\beta$}}
\newcommand{\pab}{\mbox{Pa$\beta$}}
\newcommand{\vinf}{\mbox{$v_{\infty}$}}
\newcommand{\xin}{\mbox{$\xi_{\mbox{\small N}^{4+}}$}}
\newcommand{\xic}{\mbox{$\xi_{\mbox{\small C}^{3+}}$}}
\newcommand{\xiniv}{\mbox{$\xi_{\mbox{\small N}^{3+}}$}}
\newcommand{\xiciii}{\mbox{$\xi_{\mbox{\small C}^{2+}}$}}
\newcommand{\xio}{\mbox{$\xi_{\mbox{\small O}^{5+}}$}}
\newcommand{\xisi}{\mbox{$\xi_{\mbox{\small Si}^{3+}}$}}
\newcommand{\tbl}{\mbox{$T_{\mbox{BL}}$}}
\newcommand{\hbeta}{\mbox{H$\beta$}}
\newcommand{\hdelta}{\mbox{H$\delta$}}
\newcommand{\lkha}{\mbox{LkH$\alpha$101}}
\newcommand{\heii}{He~{\scriptsize II}}
\newcommand{\hei}{He~{\scriptsize I}}
\newcommand{\hi}{H~{\scriptsize I}}
\newcommand{\civ}{C~{\scriptsize IV}}
\newcommand{\siiv}{Si~{\scriptsize IV}}
\newcommand{\nv}{N~{\scriptsize V}}
\newcommand{\niv}{N~{\scriptsize IV}}
\newcommand{\ciii}{C~{\scriptsize III}}

\title*{Massive Star Formation}
\author{Melvin G. Hoare\inst{1}\and Jos\'{e} Franco \inst{2}}
\institute{School of Physics and Astronomy, University of Leeds, Leeds, LS2 9JT, UK
\texttt{mgh@ast.leeds.ac.uk}
\and Instituto de Astronom\'ia-UNAM \texttt{pepe@astroscu.unam.mx}}
%
%
\maketitle

\section{Introduction}

The formation and evolution of massive ($>10$M\sun) stars plays a key 
role in the final fate of their parental molecular clouds, and in the 
appearance and evolution of their host galaxies. They inject large
amounts of mechanical and radiative energy creating, either by a single
star or a stellar cluster, the most spectacular gaseous nebulae in the
Cosmos. Also, they generate fast shocks which heat up the surrounding
plasma to temperatures above 10$^6$ K, and carve large interstellar 
"holes" that continuously stir the general interstellar medium. 
  
From the moment they are born their powerful outflows begin to plough 
into the surrounding molecular material. Similarly, the output of Lyman
continuum radiation, that sets them apart from lower mass stars, ionize 
the cloud and form a dense and hot H~II region. The expansion of this
photoionized nebula drives a shock wave that both compress and may trigger
further star formation on the one hand, and disperse a large portion of 
the molecular cloud on the other. As the new OB star enters the field 
population its winds and ultra-violet radiation continue to influence the
general interstellar medium. These effects strengthen after their short
main sequence lifetime, through the supergiant and Wolf-Rayet phases, and
culminates in the most energetic of stellar events, the supernova 
explosion. Thus, the energy input from massive stars, via the combined 
effects of expanding H II regions and supernova remnants, can shape the 
interstellar medium of gaseous galaxies, creating large, expanding
structures that may even vent mass and energy into the halo. It is this
litany of energetic phenomena that gives massive stars such a pivotal role
in astrophysics. Hence, it is not surprising to find that the dynamics of
these phenomena have figured strongly in John's work over the years.

\section{Accretion}

Larson \& Starrfield (1971) and John's mentor, Kahn (1974), first
investigated one of the intriguing questions in massive star formation
- how can accretion continue in the face of extreme radiation pressure.
This results from another key difference between low and high mass star
formation. The timescale for contraction in to a main sequence
configuration is shorter than the formation timescale for massive stars.
Hence, they are probably still accreting at the surface when hydrogen
burning begins in the core. As the mass increases the luminosity from 
fusion and accretion exerts a high radiation pressure on the dust grains
in the infalling cloud. In a spherically symmetric treatment Kahn (1974) 
deduced that this effect would limit the mass of a star that could be
formed by accretion to about 40 M$_{\sun}$. The adoption of more
appropriate dust parameters removes this strict limitation, but does
require very high infall rates for the ram pressure to overcome the
radiation pressure. Wolfire \& Cassinelli (1987) updated this with a full 
radiative transfer solution and concluded that normal interstellar dust 
opacities would not allow inflow to occur. Even with depleted dust models 
they found that high accretion rates ($\sim$10$^{-3}$\smpy) would be 
necessary to form the most massive stars. 

Such accretion rates were thought to be unreasonably high compared to the 
$\sim$10$^{-5}$\smpy\ expected from the collapse of a cloud initially close
to equilibrium and held up by thermal pressure alone ($\mdot_{acc} \sim 
c^{3}/G$); as is thought to be the case for low-mass star formation.  
However, several reasons have been put forward to argue for higher and
time variable accretion rates, which can also overcome the radiation
pressure problem. In any case, one might intuitively expect more massive
objects to accrete faster.

One approach taken by Norberg \& Maeder (2000) and Behrend \& Maeder
(2001) was to increase the accretion rate as the mass of the star
grows.  There is a region of the mass-luminosity plane where spherical
accretion rates can grow and keep above that needed to overcome
radiation pressure on dust in the cocoon, but below the Eddington
limit due to radiation pressure on electrons in the stellar
atmosphere. They attempted to put this on a physical basis by using
the observed relation between outflow rates and bolometric luminosity
(e.g. Churchwell 1998) and then making the accretion rate a fraction
of the outflow rate. This invokes the often found result from outflow
models that the outflow rate is a fixed fraction of the accretion
rate. Their prescription gave very high accretion rates
($\sim$10$^{-2}$\smpy) at the upper end of the mass range. However,
the observed outflow rate versus luminosity relation is likely to be
severely affected by selection effects for massive objects. Firstly,
the objects observed are likely to be at a wide range of evolutionary
stages. More importantly, Ridge \& Moore (2001) showed that when a
constant distance sample is studied there is much less of a
correlation between outflow rate and luminosity and that previous
studies were affected by Malmquist bias.

Another approach has been to use the fact that the cores that form
massive stars are not supported by thermal pressure, but by a
combination of turbulence and MHD waves (see the book edited by Franco
\& Carrami\~{n}ana 1999). Bernasconi \& Maeder (1996) used the
empirically derived Larson (1981) relations, which show the
non-thermal linewidth increasing with increasing size of cloud. This
is thought to be a natural consequence of turbulent or magnetic
support. In the simple picture in which $\mdot_{acc} \propto c^{3}$,
this gives a physical basis for the accretion rate increasing with
time as the inside-out collapse proceeds. In this way Bernasconi \&
Maeder (1996) found accretion rates up to $\sim$10$^{-4}$ \smpy\ for
the most massive stars.

McKee \& Tan (2003) have developed a turbulent core model, which results
in accretion rates about an order of magnitude higher than this ($\sim 
10^{-3}$ \smpy\ for the most massive stars). This they justify from the
observed high pressures of massive star forming cores that require higher 
linewidths than given by the usual Larson relations. They still appeal to 
the form of the Larson relations to get accretion rates increasing with 
time, although there is little direct evidence that they apply to massive 
star forming clumps at present. Indeed, the Plume et al. (1997) study upon 
which McKee \& Tan draw, states that they found no correlation between 
linewidth and size. Most of these cores also already have plenty of star 
formation activity in them, which will inevitably affect the observed
linewidths.  

The spherical accretion rates in the turbulent core model are
sufficient to overcome radiation pressure. The time variable rates
also make the star formation timescale rather independent of the mass
which helps to produce apparently coeval clusters containing a range
of masses. Their fiducial model also has a radial density distribution
on large scales that is close to r$^{-1.5}$, similar to that seen in
several studies of massive star forming regions (e.g. Hatchell \& van
der Tak 2003). This type of gradient has usually been interpreted as
being consistent with infall in a rapid star formation scenario in
which the whole region is collapsing. In the McKee \& Tan model it is
seen as a longer lived quasi-equilibrium structure supported by
turbulence.

It has been fairly clear ever since the first examples of massive
young stars were found that they form from accretion via a disc rather
than spherical infall. That is because the main manifestation of
luminous embedded sources are their ubiquitous bipolar outflows (Lada
1985). Since these most naturally arise from discs in most viable
models this has always been good indirect evidence. Accretion through
a disc is also the key to overcoming the radiation pressure in a
number of ways. Firstly, the stellar radiation is isotropic and so a
disc only intercepts a small fraction of the total
luminosity. Secondly, if the infall is concentrated through a thin
disc then the effective accretion rate is amplified. Finally, there is
likely to be a large self-shielding effect, whereby material can
accrete through the mid-plane, whilst the upper layers of the disc
take the brunt of the radiative effects.

This was investigated by Nakano (1989) and Jijina \& Adams (1996) who
showed analytically that non-spherical accretion caused by magnetised
collapse or rotation can overcome the radiation pressure for a range
of reasonable starting conditions and accretion rates. Yorke \&
Sonnhalter (2002) performed 2D axisymmetric radiation hydrodynamical
simulations on a collapsing cloud in rotation and came to similar
conclusions concerning the upper mass limit.  They find accretion
rates of about $\sim$10$^{-3}$ \smpy\ starting from a very overdense,
non-turbulent core. Their treatment of radiative transfer also showed
that a substantial amount of radiation escapes along the polar axis of
the structure where it does drive away some material, but it is along
the equatorial axis that the accretion occurs. Numerical simulations
have now moved on to three dimensions with the work of Krumholz et
al. (2005a). The application of adaptive mesh refinement (AMR)
hydrodynamical codes (see chapter by Falle) to this problem is finally
demonstrating conclusively that radiation pressure is not a barrier to
massive star formation. As material rains down on the accretion disc,
radiation bubbles repeatedly blow out perpendicular to the disc plane
and then collapse down again, but overall the accretion
continues. They also find accretion rates of order
$\sim$10$^{-3}$ \smpy\ and it would appear that a consensus may be
emerging over these kind of rates.

The effects of magnetic fields are the next major ingredient that need to 
be added to these simulations as computer power increases. It is possible 
that magnetic flows set in very early on in the accretion process as in 
the simulations of Tomisaka (2002) for the collapse of magnetised, 
rotating cores. It is unlikely that the MHD outflow mechanisms developed 
mostly for low-mass protostars can be simply scaled up to high mass 
accretors. Again the high output of UV radiation will increase the 
ionization fraction and hence, the coupling of magnetic field with the 
material. The onset of a radiative envelope (e.g. Palla \& Stahler 1992)
in the star as it adopts a more main sequence structure removes the usual 
method for the generation of strong stellar surface magnetic fields. This 
may take away the driving force in the widely used X-wind model for
protostellar outflows. Outflow mechanisms driven by magnetic fields
generated or dragged in by the disc could still operate if a strong 
stellar field is absent (Banerjee \& Pudritz 2006).  Krumholz et al. (2005b) have made an 
initial investigation of how outflows affect accretion onto massive stars 
As expected, the punching of holes along the axis by these 
flows allows the radiation pressure to be relieved in the centre, and 
accretion to continue to higher masses still.

\begin{figure}
\centering
\includegraphics[width=10cm]{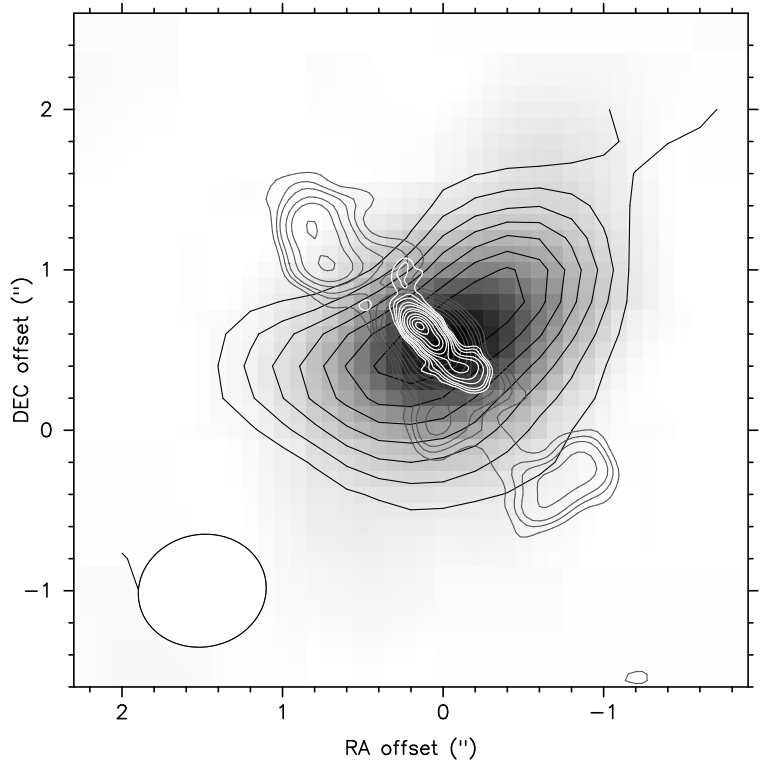}
\caption{SMA image of the 325 GHz dust emission from the massive YSO
Cep A2.  Dark contours show the integrated emission from the
CH$_{3}$CN J=18-17 transition that shows a rotational kinematic
signature consistent with a disc origin.  Grey and white contours show
the 3.6 and 1.3 cm radio continuum emission from the ionized jet in
this source.  Note the elongated structure (SE-NW) in the dust and
molecular gas perpendicular to the radio jet (SW- NE). From Patel et
al. (2005).}
\label{fig:patel}       
\end{figure}

The direct observational evidence of accretion discs around massive young 
stars is now beginning to arrive. Millimetre interferometry had delivered 
convincing observations of compact flattened structures around the more luminous 
intermediate mass young stellar objects such as GL490 (Mundy \& Adelman
1988) and G192.16-3.82 (Shepherd \& Kurtz 1999). Now sub-millimetre
interferometry has heralded the best example of a disc around a genuinely 
luminous object. Patel et al. (2005) have used the SMA to resolve an 
elongated dust and molecular structure 1000 AU in size that is perpendicular to the
radio jet in Cep A2 (Fig \ref{fig:patel}). The molecular line has a rotational 
kinematic signature which helps to confirm the disc interpretation. This 
is a taster of what ALMA will do for massive star formation studies.

We consider the alternative coalescence model by Bonnell et al. (1998)
to be unnecessary, unrealistic and unsupported. Unnecessary since
there was never really a fundamental problem with the accretion
picture as described above. The stellar densities required for it to
work are many orders of magnitude below those seen or inferred; and it
seems inconceivable that energetic events such as stellar collisions
are common without any observational sign of them.

\section{Outflows}

As mentioned above, bipolar molecular outflows have long indicated
that accretion discs are present at the heart of massive star
formation. It has been thought that outflows from massive YSOs are not
as well collimated as those from their low-mass counterparts. Much of
this impression may have arisen from low spatial resolution
single-dish molecular line observations of high-mass systems that are
on average at least ten times further away (Beuther et al. 2002a). The
strong clustering in massive star forming regions can also cause
multiple outflows to merge at low resolution. Again, the application
of interferometric observations has revealed that some regions are made
up of multiple, well-collimated outflows (Beuther et al. 2002b),
although these are not particularly high luminosity systems.

\begin{figure}
\centering
\includegraphics[width=10cm]{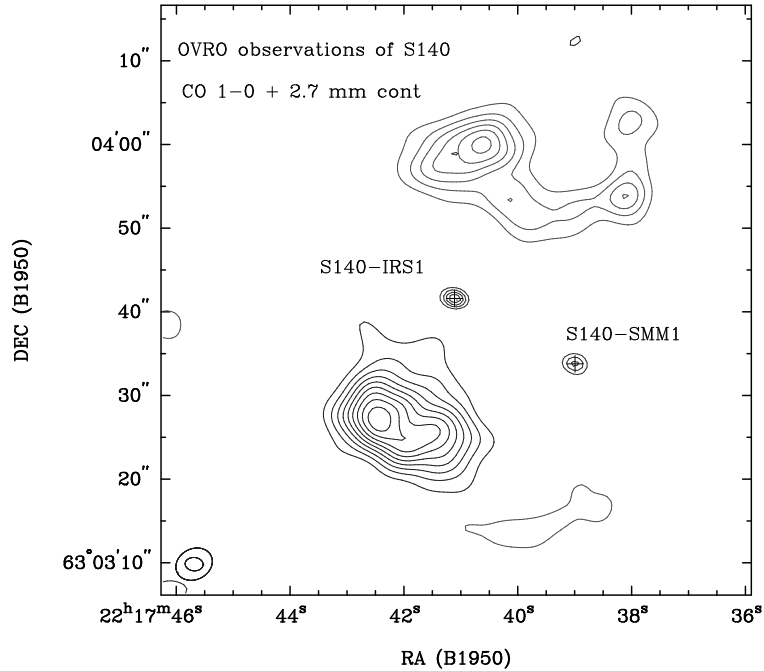}
\caption{OVRO map of the bipolar molecular outflow from S140 IRS 1. The
$^{12}$CO 1-0 observation has a resolution of about 4\arcsec\ whilst the
2.7 mm continuum emission showing the two point sources was made at a 
resolution of 2\arcsec. From Gibb, Hoare \& Shepherd, in prep.}
\label{fig:s140co}       
\end{figure}

However, other systems still appear to show little evidence for high
degrees of collimation when observed at high resolution. Fig
\ref{fig:s140co} shows OVRO observations of the outflow from S140 IRS
1. The blueshifted lobe does display an approximate bow shape, but is
not that highly collimated, whilst the redshifted lobe is severely
affected by self absorption.  If these type of outflows are driven by
jets then we should see the shock-excited emission lines
characteristic of low-mass objects. However, ground-based searches for
molecular hydrogen emission at 2$\mu$m have often turned up very
little to support this picture (Davis et al. 1998), although see Davis
et al. (2004).  Optical searches for shocked emission in the outer
reaches of molecular clouds where the flows terminate and extinction
is expected to be less have also not detected anything significant
(Alvarez \& Hoare 2005). Images from the SPITZER satellite in the
4.5$\mu$m filter are turning up numerous examples of outflow lobe
emission, most likely from rotational lines of molecular hydrogen
(Noriega-Crespo et al. 2004).  This band occurs close to where there
is a minimum in the extinction curve (Indebetouw et al. 2005) and
should be a useful probe of massive outflows.

As well as molecular outflows, the massive YSOs also have compact, ionized 
winds. These manifest themselves through thermal radio spectrum with a 
spectral index close to +0.6 expected for a constant velocity wind (Wright 
\& Barlow 1975). They also give rise to broad, single-peaked H~I 
emission lines in the IR (Bunn et al. 1995). These typically have 
FWHM of about 100 \kms\ and FWZI up to several hundred \kms\ which is a 
lower limit on the wind's terminal velocity. NLTE modelling shows that 
there is a common origin for the radio continuum and IR line emission 
(e.g. H\"{o}flich \& Wehrse 1987). The typical mass-loss rates estimated 
for these ionized winds are of order $\sim$10$^{-6}$\smpy. 

\begin{figure}
\centering
\includegraphics[bb=0 0 320 596, height=17cm,clip]{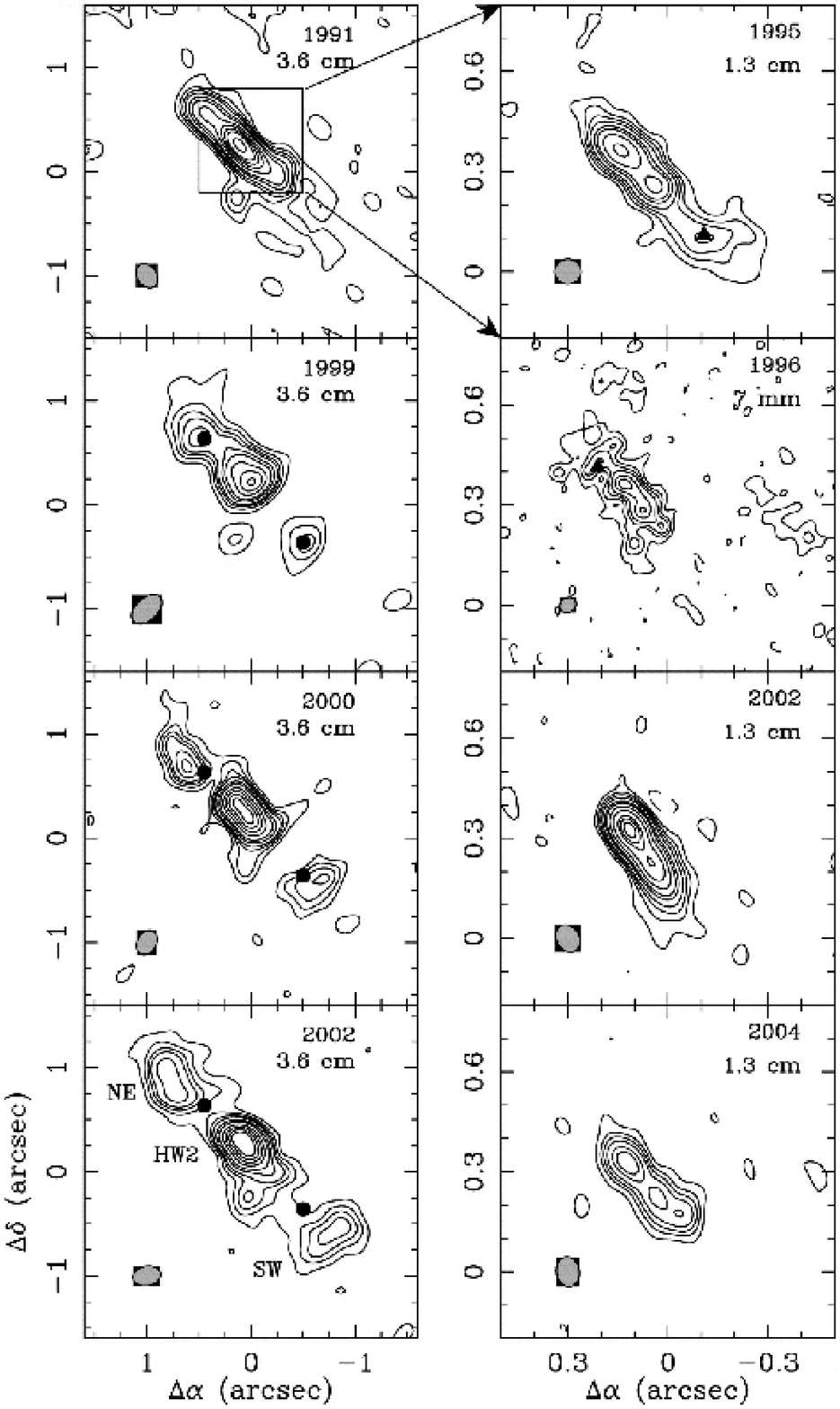}
\caption{Series of VLA 3.6 cm continuum maps of the jet from the massive 
YSO Cep A2. Note the outward motion of the knots in the jet which 
corresponds to a speed of 480 \kms\ on the sky and nearly 600 \kms\ after 
correction of inclination. The filled circles mark the position of the knots in the 1999 epoch. From 
Curiel et al. (2006).}
\label{fig:cepa2}       
\end{figure}

Clues to what is driving the ionized winds comes from resolving their
spatial structure. Here the picture is also mixed as it is for the
molecular outflows (see review by Hoare 2002). In some cases radio
jets are seen as in their low-mass counterparts.  By far the most
spectacular example is the 2.6 pc long wiggling jet seen in the GGD27
system (Mart\'{i} et al. 1993). This powers radio and optical
Herbig-Haro objects at its termination and looks just like a scaled up
low-mass system. Proper motion studies reveal velocities of 500 \kms\
(Mart\'{i} et al. 1998). Other examples are G35.2N (Gibb et al. 2003) and
IRAS (Rodr\'{i}guez et al. 2005) and possibly W3 IRS 5d2
(Wilson et al 2003), which are mostly much more knotty and have a more
exaggerated point symmetry.  In other cases we just resolve the base of
the jet; the best example being Cep A2 as in Fig \ref {fig:patel}, and
GL 2591 also falls in this category (Trinidad et al. 2003).  The fast
motion in the Cep A2 jet has recently been confirmed by proper motion
studies (Curiel et al. 2006) (Figure \ref{fig:cepa2}). One expects
that such highly collimated flows are driven by MHD mechanisms in the
star-disc system as for low-mass outflows.

However, jets are not the only radio morphology seen in massive YSOs. The
exciting source of the peculiar bipolar H~II region S106, which otherwise 
shows all the characteristics of a massive YSO (Drew et al. 1993), has 
radio emission elongated perpendicular to the outflow axis (Hoare et al. 1994; Hoare \& Muxlow 
1996; Hoare 2002). S140 IRS 1 also has such an 
equatorial wind of about 500 AU in size (Hoare \& Muxlow 1996; Hoare 
2002). Originally interpreted as a jet (Schwartz 1989; Tofani et al. 1995) 
it is clearly perpendicular to the large scale bipolar molecular outflow 
and to the monopolar IR reflection nebula at the base of the blueshifted 
outflow cavity revealed in speckle observations (Schertl et al. 2000; Alvarez 
et al. 2004). Proper motions confirm that this structure is not moving 
outwards as would be expected for a jet (Hoare 2002). Another possible 
example of an equatorial wind is in GL490 (Campbell et al. 1986), although 
the extension in the radio emission along the disc plane (Mundy \& Adelman 
1988) is somewhat noisy.

To drive such equatorial winds it has been proposed by Drew et
al. (1998) that the radiation pressure due to central star and inner
disc act on the ionized gas at the surface of the disc. The gas is
pushed sideways, across the surface of the disc forming an equatorial
wind. This is the same mechanism that drives the stellar winds in
field main sequence stars where the terminal velocity is a few
thousand \kms, i.e.  of order the escape speed at the stellar surface
(Prinja et al. 1990). In the massive YSO case, the surface gravity in
the disc atmosphere a few stellar radii out from the star is lower and
so the typical speed of the equatorial wind is a few hundred
\kms. Hence, this mechanism has the potential to explain both the
morphology, speed and mass-loss rate of the observed equatorial wind
systems. Initial attempts to simulate the line profiles arising from
such winds have too high a rotational component and result in
double-peaked line profiles whereas only single-peaks are usually seen
(Sim et al.  2005).

Note that this radiatively pressure driven flow is very different from
the photo-evaporation mechanism discussed by Franco et al. (1989) and
Hollenbach et al. (1994). That drives a much slower, thermal flow from the 
outer regions of the disc and they did not consider the effect of 
radiation pressure on the inner disc. Both mechanisms could operate 
simultaneously, but if the accretion disc does extend down close to the 
star there is little doubt that the radiation pressure acting on the gas 
will drive such a wind and is likely to dominate the dynamics.

\section{The Transition to UCHII Region}

It has long been a puzzle as to why the massive YSOs do not ionize their 
surrounding molecular gas as soon as they become luminous enough. All 
objects above about 10$^{4}$L\sun\ should emit copious amounts of Lyman 
continuum radiation if they are fully on the zero age main sequence 
(ZAMS). It is commonly accepted that massive stars begin core hydrogen 
burning whilst still accreting and so the ZAMS assumption appears 
justified. 

One possibility is that the infall of molecular material 'quenches'
the H~II region (Walmsley 1995). The critical accretion rate for this
to occur is obtained from inserting an infalling spherically symmetric
density distribution into the Str\"{o}mgren radius equation. The rates
are high, but not as high as the $\sim$10$^{-3}$\smpy\ needed to
overcome radiation pressure on the dust in the same spherical
treatment. In this scenario, the H~II region is not absent, but just
very dense ($\sim10^{12}$~cm$^{-3}$ with a Str\"{o}mgren radius close
to the star. One problem with this approach is that the infall would
have to be approximately spherically symmetric in order to stop the H
II region breaking out in low density directions. In any accretion
scenario, most of the infalling material will arrive on the disc
rather than the star and a bipolar H~II region would ensue with
ionized lobes above and below the disc (Franco et al. 1989).  It also
neglects the effect of winds and outflows which are likely to push the
inner radius of the infalling envelope some distance away from the
star, again mostly likely in the polar directions.

Another picture was developed by Tan and McKee (2003) in which strong
outflow rather than infall confines the H~II region. They used an
approximate X-wind outflow density distribution where the H~II region
propagates along the cavity along the axis. These models are basically
X-wind jets with extra photo-ionisation. They predict ionized zones
that are very narrow (few AU), whereas most high-mass radio jets are
resolved across the minor axis ($\sim$ 50 AU) (Mart\'{i} et al. 1999;
Curiel et al. 2006). Jets from high-mass objects are also likely to be
significantly ionized as they are launched. Low-mass jets are
partially ionized, and is difficult to predict at this moment how the
ionization fraction of MHD driven jets would change for higher mass
stars. If they are significantly ionized then their ability to confine
the H~II regions would be dramatically reduced.

Again a difficulty is that as soon as the central star becomes at all
luminous it will start to drive strong stellar winds due to radiation
pressure. These will open up cavities further exposing more material to 
the ionizing radiation. The equatorial winds discussed above would also 
not be able to confine the H~II region since they do not cover the polar 
regions. Indeed, in the models by Drew et al. (1998) the polar regions 
are occupied by a normal O star stellar wind flowing at a few thousand 
\kms. 

\begin{figure}
\centering
\includegraphics[width=10cm]{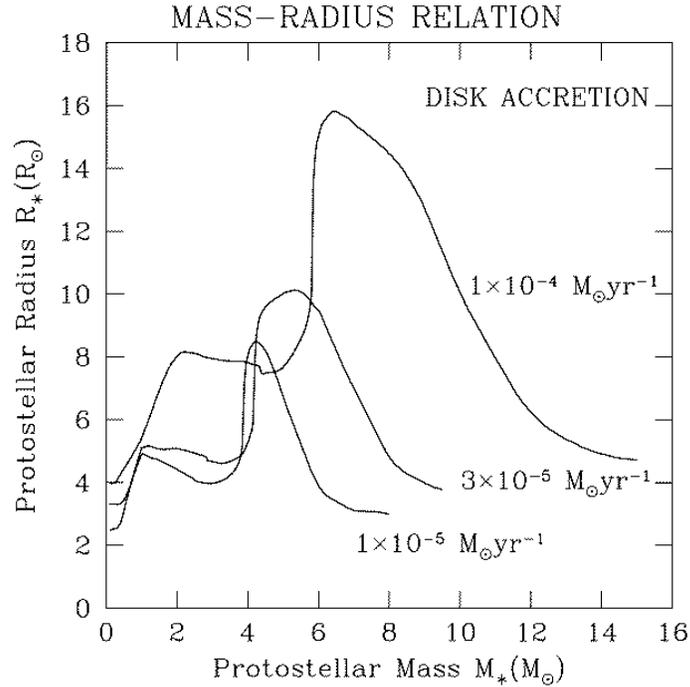}
\caption{Variation of the radius of an accreting star as it grows in mass
with the steady accretion rates shown. From Palla \& Stahler (1992).}
\label{fig:palla}       
\end{figure}

One possible solution is to tackle the problem at source by examining 
whether the star really is emitting large amounts of Lyman continuum
radiation during the massive YSO phase. Although the core is very likely 
to be already on the main sequence it is unclear that the outer layers of 
the star have contracted fully into a main sequence configuration. It is 
well known that accreting stars swell up well beyond their ZAMS radius, 
mainly due to shell deuterium burning (Palla \& Stahler 1992). What is 
more, the higher the accretion rate the more the radius increases and 
extends to higher masses before the rapid contraction onto the main 
sequence occurs. The calculations by Palla \& Stahler only consider 
accretion rates up to $\sim$10$^{-4}$ \smpy\ and masses up to 15 M\sun\ 
(see Figure \ref{fig:palla}), but the current thinking points towards 
accretion rates as high as $\sim$10$^{-3}$ \smpy. Inspection of Figure \ref
{fig:palla} shows that we would then expect a much greater swelling, up to 
around 30 R\sun, before the contraction to the main sequence at around 30~M
\sun\ (L$\sim 10^{5}$ L\sun) for these rates of accretion (Palla, private 
communication).

The key point here is that a larger radius means lower effective
temperatures. If ongoing accretion keeps the effective temperature below 
about 30~000~K then there is no need to invoke any mechanism to quench the 
H~II region since there will not be a significant Lyman continuum output 
from the star in any case. Future accretion models need to be calculated 
with a fully self-consistent treatment of the stellar mass-radius relation 
to investigate this aspect. The McKee \& Tan (2003) models, in which the 
stars accrete at rates significantly above $\sim$10$^{-4}$ \smpy, implement 
an approximate treatment of the the Palla \& Stahler (1992) results and 
their stars join the main sequence at masses of around 20 M\sun\ as 
expected. However, their radii never get above the 15 R\sun\ for Palla \& 
Stahler's $\sim$10$^{-4}$ \smpy\ case suggesting that these have been
under-estimated for the highest accretion rates. The Behrend \& Maeder
(2001) mass-radius relation shows a similar pattern of swelling towards 
larger radii, but again not as much as one would expect based on Figure 
\ref{fig:palla}, since their accretion rates exceed $\sim$10$^{-3}$ \smpy\ 
through the relevant mass range.

Nakano et al. (2000) used a polytropic model to estimate that a young star 
accreting at $\sim$10$^{-2}$ \smpy\ would have a maximum radius of about 30 
R\sun. In a different context Kippenhahn \& Meyer-Hofmeister (1977) 
calculated the affect on the radii of stars already fully on the main 
sequence of accretion at rates exactly in the $\sim$10$^{-4}$ \smpy\ to 
$\sim$10$^{-2}$ \smpy\ range of interest here. They found that the stars can 
expand greatly, e.g. about 100 R\sun, occupying the supergiant part of the 
Hertzsprung-Russell diagram before returning to the main sequence at 
higher mass. This was not because of deuterium burning, but simply because 
the accretion timescale ($M/\mdot$) is shorter than the thermal adjustment 
or Kelvin-Helmholz timescale. The expanded phase lasts for the accretion 
timescale if the accretion rate is constant, which is 10$^{4}$ years for a 
10~M\sun\ star accreting at $\sim$10$^{-3}$ \smpy. If the accretion rate is 
increasing with time this phase would be longer still.

Overall, if the accretion rates during the hot core/massive YSO phases are 
as high as $\sim$10$^{-3}$ \smpy, then it is plausible that the reason for 
the lack of an H~II region is simply that ongoing accretion keeps the 
effective temperature of the star too low. This scenario would also 
explain several other features of massive YSOs. If the ongoing accretion 
keeps the stellar radius high then it also lowers the surface gravity. It 
has been noted several times in the literature how the spectroscopic 
characteristics of massive YSOs resemble those of evolved OB stars (Simon 
\& Cassar 1984). Most of the massive YSOs amenable to near-IR spectroscopy 
have luminosities around a few 10$^{4}$ L\sun\ and thus are about 12 M\sun, 
just where accretion could easily result in a low surface gravity. The 
emission line spectrum with broad profiles of a few hundred \kms\ are 
reminiscent of B supergiants. The low surface gravity leads directly to 
low escape speeds and increased mass-loss. 

Another feature of large central stars would be to help explain the near-
IR line profiles in equatorial wind sources like S106IR. The current 
attempts to model the H I lines result in double-peaked rather than 
single-peaked lines (Sim et al. 2005).  If the central star is larger than 
the ZAMS one assumed, then the rotation of the inner disc is much slower, 
reducing the rotational splitting of the disc wind. The slower, denser 
stellar wind from the poles may also help fill in the profile to improve 
agreement with observations. Such a wind could also help trap the Lyman 
continuum radiation before it can ionize the surroundings. 

Whichever of the above mechanisms prevents the formation of an H~II region 
during the massive YSO phase; quenching by infall, outflows or accretion 
swelling the star, all are associated with ongoing accretion. This would tend to indicate that the 
onset of the H~II region phase is not to do with 
the star evolving to a stage where it is hot enough to ionize hydrogen, 
but more to do with the cessation of accretion. If accretion swelling the 
star is the cause, then the end of the high accretion rate phase will then 
allow the star to contract down onto the ZAMS. In so doing, the Lyman 
continuum output will increase and the H~II region phase begin.

When ionization of the surrounding material does begin the ionization
front will move faster in the low density directions. These are most
likely to be the polar lobes due to the flattening effect of the
centrifugal barrier during the infall phase and bipolar outflows
during the YSO phase. In this scenario, one would expect the youngest,
most compact H~II regions to be bipolar. Within the class of so-called
hyper-compact H~II regions (HCHII, see Kurtz 2005) there are examples
of bipolar objects. These also tend to display broad recombination
lines (\gtappeq 40 \kms), which has been used as another criterion to
set some of the hyper-compact objects apart as a new class (Jaffe \&
Martin-Pintado 1999; Sewilo et al 2004).

\begin{figure}
\centering
\includegraphics[width=10cm]{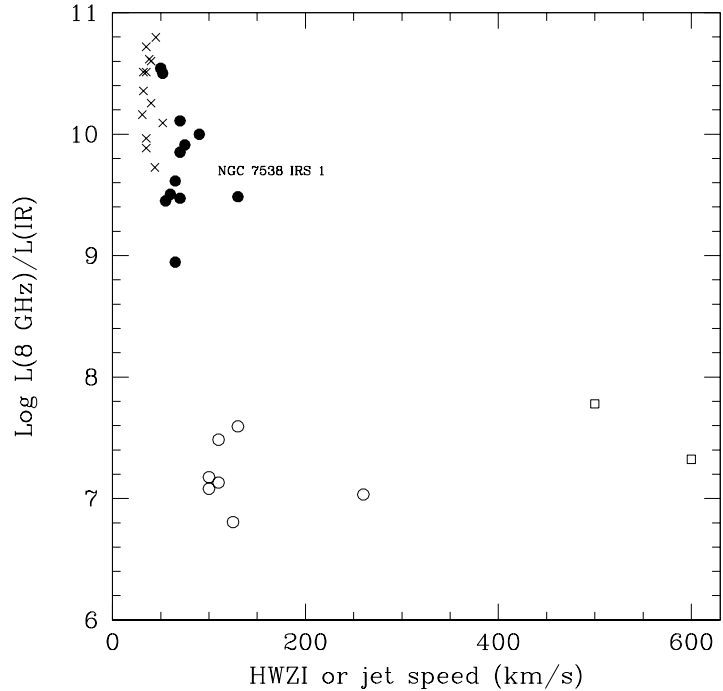}
\caption{Ratio of the radio luminosity at 8~GHz (W~Hz$^{-1}$) to the 
bolometric luminosity from the IR (L$_{\sun}$) for UCHII regions 
(crosses), HCHII regions (solid circles) and massive young stellar object 
wind sources (open circles) and jets (open squares).  Line widths are 
HWZI.}
\label{fig:ratio}       
\end{figure}

Figure \ref{fig:ratio} shows a plot of radio luminosity versus linewidth for
UCHII regions, HCHII regions, young stellar wind sources, and jets
(see Hoare et al. 2006 source of these data). Unlike in Hoare et al. (2006), here we
have plotted the half-width-zero-intensity (HWZI) for the
recombination lines rather than FWHM. For the UCHII and hyper-compact
H~II regions this does not make a great deal of difference, but some
stellar wind sources appear to have a narrow optically thin component
which dominates the FWHM whilst a broader optically thick component
dominates the HWZI. In these cases the HWZI gives a better indication
of the kind of speeds the wind is attaining. However, it is a more
difficult parameter to work with since where the line returns to the
continuum is dependent on the signal-to- noise of the spectrum. We
have also added the two massive YSO jet sources that have measured
proper motions to this plot: GGD 27 and Cep A2. These have been
corrected for the inclination where known and thus measures the full
speed of the gas. It is unfortunate at present that the jet sources
are not amenable to IR spectroscopy to probe their line profiles as
they are deeply embedded and not directly visible.

In Figure \ref{fig:ratio}, NGC 7538 IRS 1 appears to stand out
relative to the UCHII regions and stellar wind/jet sources. There is
even a question mark over this source since the Sewilo et al. (2004)
measurement plotted is for a low frequency line that may be affected
by pressure broadening and NLTE effects, whilst the mm-wave
recombination lines show HWZIs nearer 80 \kms\ (Jaffe \&
Martin-Pintado 1999). Other designated hyper-compacts appear to form
more of a continuum with the UCHII regions in this plot and in the
size-linewidth relation (Hoare et al. 2006). There is no real break
around 40-60 \kms. NGC 7538 IRS 1 is distinctive in that the bipolar
lobes have velocity widths in excess of 100 \kms, rather like the
massive YSO stellar wind sources, but it is much more luminous in the
radio continuum.  The brightness of the radio lobes has recently been
shown to be getting significantly fainter on a timescale of 10 years
(Franco-Hern\'{a}ndez et al. 2004). The short lifetime for this phase
would be consistent with the rarity of objects like NGC 7538 IRS 1.

\begin{figure}
\centering
\includegraphics[width=10cm]{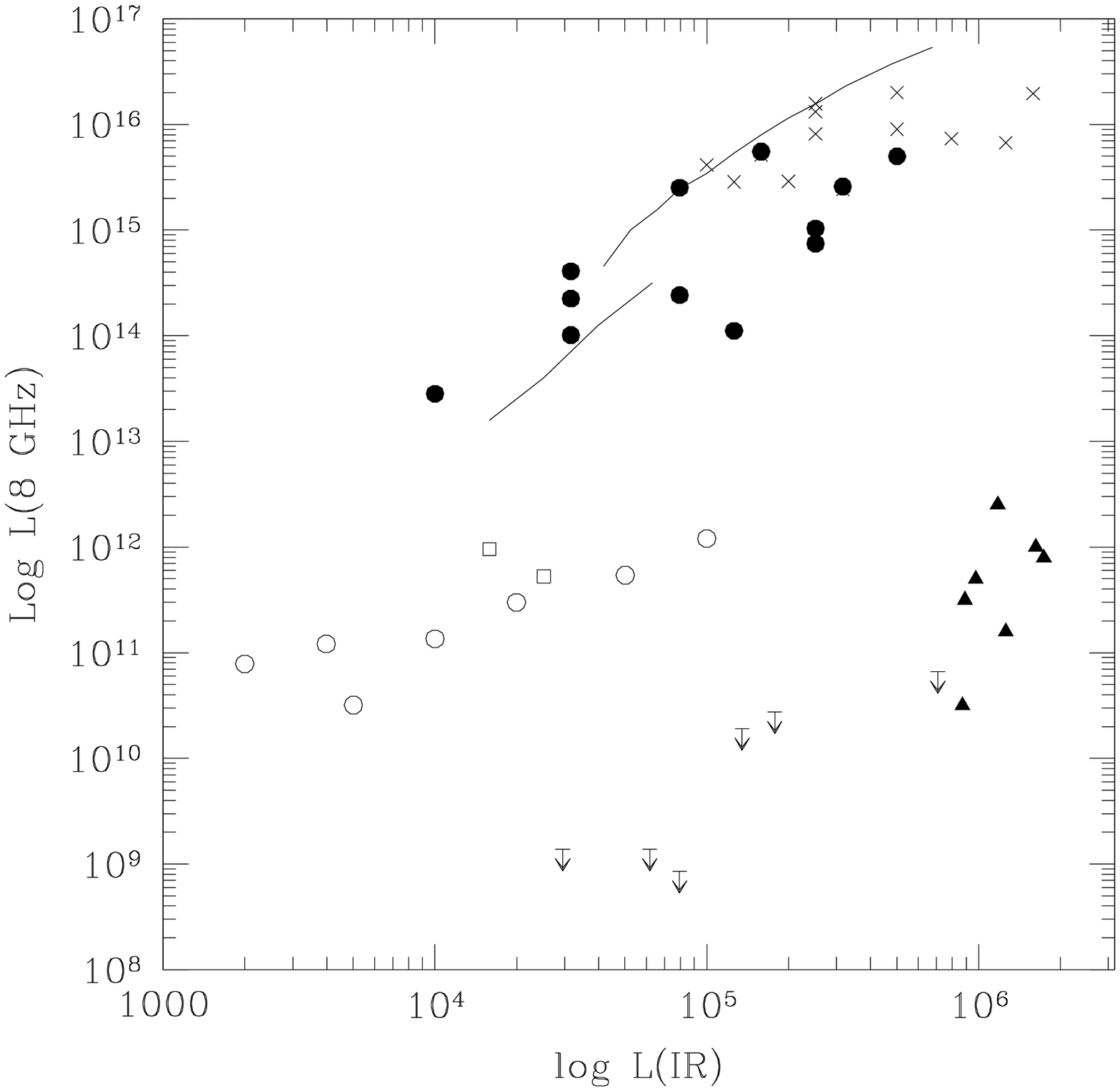}
\caption{Plot of the radio luminosity at 8~GHz (W~Hz$^{-1}$) versus
the bolometric luminosity from the IR (L$_{\sun}$) for UCHII regions
(crosses), HCHII regions (solid circles) and massive YSO wind sources
(open circles) and jets (open squares), evolved OB stars (filled
triangles) and MS OB stars (upper limits). OB star data from thermal
emitters in Bieging et al. (1989).  Solid lines represent the expected
optically thin radio luminosity at 8 GHz from MS stars at the given
luminosity using O star parameters from Martins et al. (2005) and B
star parameters from Smith et al. (2002).}
\label{fig:radlum}       
\end{figure}

Another object commonly placed in this category is MWC 349A.  However,
its linewidths are somewhat narrower and the mm-wave recombination
lines are masing which is not seen in other hyper-compacts
(Mart\'{i}n-Pintado et al. 1989). There are still question marks over
the evolutionary stage of MWC 349A and even whether it is a young or
post-main sequence object (e.g. Meyer et al. 2002).

It is tempting to associate these bipolar hyper-compacts with the
turn-on phase of the H~II region as it expands down the previously
excavated bipolar outflow cavities. For simple H~II region expansion
it is possible for the gas to get up to a few times the ionized gas
sound speed (10~\kms) when travelling down the steep density gradients
expected away from the disk plane. Franco et al. (1989, 1990) found
speeds up to around 30 \kms\ for steep power-law density gradients
and for some disk-like density gradients.

In particular, the bipolar H~II regions resulting from disk-like density 
distributions have some interesting features that may be relevant to 
features appearing at the density distributions expected at the turn-on 
phase, and need further exploration. For instance, as pointed out by 
Franco et al (1989), the dynamics of the H~II regions depends on the 
details of the density distribution, and its appearance will be affected
by recombination fronts that can result in bipolar molecular outflows, originally generated by the 
photoionized plasma. This type of phenomenon
and the fragmentation created by instabilities in the ionization-shock 
fronts (Garcia-Segura and Franco 1996), along with complex structure of 
combined ionization-photo-dissociation fronts (Diaz-Miller et al 1998), may 
provide some hints to the origin of the multiple flows with knotty 
structures discussed above.

Another model which is commonly invoked to explain bipolar
hyper-compacts, is the photo-evaporating disk model by Hollenbach et
al. (1994) mentioned above. This is a variation of the Franco et al
(1989) model, in which the disk is now supposed to be formed from the
gas accreted into the massive protostar. Lugo et al. (2005) show that
such a picture can match the radio continuum spectrum for MWC 349A and
NGC 7538 IRS 1. Such a thermal expansion model is capable of
explaining the relatively low velocities in MWC 349A, but in no way
can it account for the much faster flows in NGC 7538 IRS 1. MWC 349A
does not appear to be deeply embedded in a molecular cloud and so a
circumstellar disc origin is more likely. However, NGC 7538 IRS
1 is deeply embedded with evidence of large
and small scale outflow activity (Davies et al. 1998; Kraus et al. 2006).

To explain the velocities of order hundred \kms\ in NGC 7538 IRS 1 and its 
intermediate nature between wind/jet sources and UCHII regions it is 
natural to think of some kind of transition object. One scenario is that a 
fast stellar wind, perhaps more like a main sequence radiatively driven 
wind travelling at around a thousand \kms\ is beginning to blow down the 
bipolar cavity and entrain material. As the star contracts onto the main 
sequence, as it must do in order to produce the Lyman continuum in the 
first place, such a wind will inevitably come with the increase in extreme 
UV radiation. It is certainly within the realms of possibility that such a 
wind will get mass-loaded with material from the walls of the bipolar
cavity.

In this picture one would also expect to find much larger bipolar objects 
as the wind and ionization break out the axes. One such object is the 
bipolar H~II region S106. The ionized lobes of this object have broad 
lines with HWZI=95 \kms\ (Solf \& Carsenty 1982; Jaffe \& Martin-Pintado 
1999) and the full extent is about 0.7 pc.  It has a limb-brightened 
appearance (Felli et al. 1984), which together with the presence of actual 
line-splitting indicates a swept-up bipolar shell structure. With his 
usual insight Dyson (1983) developed a model for S106 whereby a star turns 
on in a plane-stratified medium where the H~II region quickly breaks out 
in the polar direction and is followed by the stellar wind sweeping the 
nebula into a thin bi-cylindrical structure. At the ends of the nebula 
the shell is predicted to be Rayleigh-Taylor unstable and break up. Such a 
picture is consistent with the latest dramatic near-IR pictures of the 
nebula by 8 m telescopes (Oasa et al. 2006). To get the right expansion
velocity of the shell Dyson's model used main sequence parameters for the 
stellar wind, which are consistent with its overall luminosity and 
spectral type. However, the spectral characteristics of the exciting 
source are more like those of a massive YSO (Drew et al. 1993). It is also 
one of the equatorial wind sources (Hoare et al. 1994), although this does 
not preclude a main sequence wind blowing from the poles of the star.

Objects like S106 are rare, in fact as an optically visible example it
is currently unique. Dyson (1983) rightly pointed out that these
objects would be rare with only a few in the whole galaxy. K3-50A
(Turner \& Welch 1984; De Pree et al. 1994) and W49N A2 (De Pree et
al. 2004) appear to be similar objects, although they have strong
ionized emission in the disc plane, whereas S106 has a lack of
emission in the disc plane. NGC 6334 A also shows bipolar lobes with a
significant velocity gradient, but this time breaking out from a more
shell-like central source (De Pree et al. 1995). G5.89-0.39 may be
another example of this structure (Acord et al. 1998). The dynamical
ages of these objects also appear very short, again attesting to their
rarity, but it also seems that the vast majority of ultra-compact H~II
regions do not have a bipolar morphology.

\section{Ultra-compact H~II Regions}

A global review of the most important features for the evolution of
HII regions, from the ultra compact stages to the extended phases, is
given by Dyson and Franco (2001). The usual definition of UCHII
regions is that they have sizes smaller than 0.1 pc and electron
densities above to 10$^4$ cm$^{-3}$, and are located in the inner,
high-pressure, parts of the parental molecular clouds (see Kurtz and
Franco 2002, Kurtz et al. 2000 and Hoare et al. 2006 for reviews). As
stated above, there is also an even more compact stage referred to as
the hyper-compact phase.

There are about a thousand candidates identified, and they are classified 
into a variety of morphologies, ranging from spherical to irregular. Their 
internal density structure can be derived with the method of spectral 
index analysis, and power-law decreasing distributions with steep slopes
(exponents below -2) have been obtained for some objects (Franco et al 
2000). The large number of objects at this UC stage may be due to two
independent factors; one is that some of them have extended emission that
was missed due to selection effects in the early studies (Kurtz et al.
1999; Kim and Koo 1996, 2001), and the other one is that they do not seem
to be the short-lived phase predicted by simple models of expanding H~II 
regions. The first factor simply implies that the actual number of UCHII
regions has to be revised, and is probably smaller than initially stated. 
In addition, the possibility of having UCHII regions with a compact core 
and a large, more diffuse, envelope seems to be a logical consequence of
the steep density gradients found by Franco et al. (2000). The second 
factor implies that the simple model is not directly applicable at this UC 
stage, and there may be several mechanisms that would confine them and 
delay their expansion. Indeed, there are a number of models proposed to 
explain the apparently slower growth rates, including the possibility of 
reaching pressure equilibrium inside cloud cores.

The cores of massive molecular clouds are clumpy, turbulent, and highly
pressurised regions. Their total central pressures are above 10$^6$ dyn 
cm$^{-2}$ (see Garcia-Segura and Franco 1996), and some H~II regions can 
reach pressure equilibrium within the central uniform-density core. If this
occurs, the resulting pressure-confined regions will have sizes of $2.9
\times 10^{-2} F^{1/3}_{48} T^{2/3}_{4} P^{-2/3}_{7}$ pc, and densities around 
$3.6 \times 10^{4} P_{7} T^{-1}_{4}$ cm$^{-3}$, where the UV photon flux is 
$F_{48} = F/10^{48}$ photons s$^{-1}$, $P_{7} =P/10^{7}$
dyn cm$^{-2}$, and the photoionized plasma temperature is $T_{4} =T/10^{4}$ K 
(the sizes are similar for the case of massive star wind-driven bubbles; see
Garcia-Segura and Franco 1996, and Kurtz et al 2001). As indicated above, 
these sizes and densities are already typical of UCHII regions. If one 
includes dust absorption, the sizes are shortened even more (see Diaz-
Miller et al 1998 and Arthur et al. 2004), making this mechanism even more 
attractive to confine the growth of UCHIIs. These, however, are only a few parts 
of a much more complicated story; the simplified pressure equilibrium
scheme does not include the presence of clumps in highly dynamical cloud
cores, nor the motion of stars inside the cloud. In addition, it does not 
explain the variety of morphologies that are already known for these objects.

Again, John's physical intuition was used to explore the effect of
these cloud core clumps in the expansion of the ionization front. The
net result is that, as the clumps are photo-evaporated inside the H~II
region, the growth of the ionized region stalls because the
recombination rate increases as the mass and density of the
photoionized gas is increased (Dyson et al. 1995; Williams et al 1996;
Redman et al. 1996; Redman et al. 1998). Thus, the destruction of
clumps is an effective mass-loading mechanism that reduces the
expansion rate of the H~II region, leading to a substantially longer
lifetime of the ultra compact stage.

Another issue that has been considered in some detail is the origin of
cometary UCHIIs. Their structure, with a bright head and a more
diffuse extension on one side, resembles the morphological features of
comets.  Also, cometary UCHII regions are among the most common
objects (they represent at least 20 \% of them and maybe many more
(Hoare et al. 2006)), and their shapes have been ascribed to either
bow-shocks generated by the motion of the exciting star (see Van Buren
and McCray 1988 and Van Buren and Mac Low 1992) or by the presence of
a density gradient on one side of the parental cloud, that creates a
"champagne flow" (e.g., Tenorio-Tagle 1979). A recent study by Arthur
and Hoare (2006), which modelled the structuring and emission of the
objects created by both types of models, indicates that the most
likely origin is due to density gradients with a champagne flow. 
In these numerical models the affect of stellar winds was included
which swept the H~II region into a thin shell and results in
the limb-brightened appearance of many cometaries. Once again
John had considered such a problem many years before whereby a
stellar wind bubble expanded in an H~II region with a density
gradient (Dyson 1977).

The motion of stars have a role in the structuring of both H~II regions and
clouds. The fact that the star moves back and forth from high to low 
densities, as it moves within the gravitational field of the cloud, a wide
variety of transient and complex structures are created (see Franco et al
2006).

\section{The impact of massive stars}

The birth of a massive star  is a strong source of energy that excites, stirs and 
ionizes the parental molecular cloud. The UV photon flux results in expanding 
H~II regions, while the strong winds create expanding bubbles that are internal
to the main body of the H~II region.  In addition, the non-ionizing UV radiation 
is able to dissociate the molecular gas creating a layer of atomic hydrogen that 
envelops the H~II region. Thus, several distinct regions form around the young 
massive star, creating a nested structure with a wind-driven bubble, a photoionized 
H~II region, and a photo-dissociated (PDR) region. The details, extent, and structure 
of these regions depend on the mass of the young star, the ambient density
structure, and the age of the star. Regardless of these details, however, the
combined action of radiation and winds evaporates the gas surrounding the new
stars and finally destroys the molecular cloud, effectively shutting-off the star 
formation  process. The evaporation of the cloud occurs via peripheral blisters,
or champagne flows.

As discussed by Franco et al (1994) and Diaz-Miller et al (1998), then, the 
star-forming capacity of molecular clouds is limited by cloud destruction from 
massive stars. The limit on the number of massive stars is set by the overlap of 
internal H~II and PDR regions. Thus, one of the main roles played by the energy
injection from massive stars is to set a star formation control that regulates the
number of stars formed within any star-forming cloud. This, in turn, implies
that such a process also regulates the actual star-forming cycle of galactic systems.

After the parental cloud is destroyed, the expanding shells continue their evolution
in the general interstellar medium, creating supershells. Most of the supernovae 
resulting from the newly formed stellar group actually explode inside these cavities.
Thus, at the late stages of massive star lives, the impact is shaping and stirring
the general ISM.

%
%
%



\end{document}